\documentclass{ecai}
\usepackage[hyphens]{url}  
\usepackage{times}
\usepackage{graphicx}
\usepackage{latexsym}
\newcommand{\floor}[1]{\lfloor #1 \rfloor}

\newcommand{\round}[1]{\ensuremath{\lfloor#1\rceil}}

\begin{document}

\title{Yet Another Comparison of SAT Encodings for the At-Most-K Constraint}

\author{Neng-Fa Zhou\institute{CUNY Brooklyn College \& Graduate Center, email: zhou@sci.brooklyn.cuny.}}

\maketitle
\bibliographystyle{ecai}

\begin{abstract}
  The at-most-k constraint is ubiquitous in combinatorial problems, and numerous SAT encodings are available for the constraint. Prior experiments have shown the competitiveness of the sequential-counter encoding for k $>$ 1, and have excluded the parallel-counter encoding, which is more compact that the binary-adder encoding, from consideration due to its incapability of enforcing arc consistency through unit propagation. This paper presents an experiment that shows astounding performance of the binary-adder encoding for the at-most-k constraint.
\end{abstract}

\section{INTRODUCTION}
The at-most-k constraint, which is denoted as $\le_k(x_1,x_2,\ldots,x_n)$ in this paper, commonly occurs in combinatorial problems. Numerous encodings have been proposed for the constraint. Prior experiments \cite{Frisch10,Wynn18} have shown that Sinz's \textit{sequential-counter} (SC) encoding \cite{Sinz05}, which uses unary presentation for counters, is competitive for $k > 1$. SC introduces $O(nk)$ new variables, and generates $O(nk)$ clauses. This encoding is not scalable for large $n$ and $k$. As a sequel, Sinz also proposed a \textit{parallel-counter} (PC) encoding, which uses binary representation for counters, and is more compact than SC. PC uses incomplete adders that only propagate 1's.  The \textit{binary-adder} (BA) encoding, which uses complete adders capable of propagating both 0's and 1's, has been shown to be competitive for encoding integer-domain variables and arithmetic constraints \cite{ZhouK17}. The at-most-k constraint can be treated as a special linear constraint. One question arises: how competitive is the binary-adder encoding for the at-most-k constraint?

This paper addresses the above question. It presents an experiment comparing the BA, PC, and SC encodings. As the benchmark used in the experiment, the \textit{pigeonhole} problem, includes the at-most-one constraint, this paper also surveys SAT encodings for the at-most-one constraint, and presents comparison results of these encodings.

\section{SAT ENCODINGS for AT-MOST-K}
This section gives an overview of each of the SAT encodings for the at-most-k constraint used in the experiment. 

The \textit{\textbf{ pairwise}} (PW) encoding for $\le_1(x_1,x_2,\ldots,x_n)$ decomposes the constraint into $\neg x_i \lor \neg x_j$, for $i \in 1..n-1$ and $j \in i+1..n$. PW introduces no new variables. However, it generates $O(n^2)$ clauses, and is therefore not viable for large $n$. PW is utilized to handle the base case when $n \le 4$ in the recursive encodings.

The \textit{\textbf{ bisect}} (BS) encoding  for $\le_1(x_1,x_2,\ldots,x_n)$ splits the variables into two groups $G_1 = \{x_1,x_2,\ldots,x_m\}$ and $G_2 = \{x_{m+1},\ldots,x_n\}$ when $n > 4$, where $m = \floor{\frac{n}{2}}$. It introduces a new variable $b$ as the commander variable for $G_1$, and uses $\neg b$ as the commander for $G_2$. BS decomposes the constraint into the following:
\begin{tabbing}
  aaa \= aaa \= aaa \= aaa \= aaa \= aaa \= aaa \= aaa \= aaa \= aaa \= aaa \= aaa \= aaa \= aaa \kill
\>   (BS-1)\> \> For  $i \in 1..m$: $x_i \Rightarrow b$ \\
\>   (BS-2)\> \> For  $i \in m+1..n$: $x_i \Rightarrow \neg b$ \\  
\>   (BS-3)\> \>   $\le_1(x_1,x_2,\ldots,x_m)$ \\
\>   (BS-4)\> \>   $\le_1(x_{m+1},\ldots,x_n)$   
\end{tabbing}
Constraint BS-1 forces $b$ to be 1 if any of the variables in $G_1$ is 1. Constraint BS-2 forces $b$ to be 0 if any of the variables in $G_2$ is 1. Since $b$ cannot be both 0 and 1 at the same time, it is impossible for one variable in $G_1$ and another variable in $G_2$ to be 1 simultaneously. Constraints BS-3 and BS-4 recursively enforce at-most-one on the two groups. The BS is a special case of the \textit{bimander encoding} \cite{Nguyen15}, which generalizes the \textit{binary encoding} \cite{FrischPDN05} and the \textit{commander encoding}\cite{Klieber07}. The number of clauses generated by BS is $O(n\log_2(n))$, and the number of new variables introduced is $O(n)$.

The \textit{\textbf{ product}} (PD) encoding \cite{JChen10} for $\le_1(x_1,x_2,\ldots,x_n)$ arranges the variables on an $m\times m$ matrix $M$ when $n > 4$, where $m = \sqrt n$. It introduces two vectors of new variables $<$$u_1, u_2, \ldots, u_m$$>$ and $<$$v_1, v_2, \ldots, v_m$$>$, where $u_i$ represents row $i$ and $v_j$ represents column $j$. In case $n$ is not a square number, the extra entries of $M$ are filled with 0. PD decomposes the constraint into the following:
\begin{tabbing}
  aaa \= aaa \= aaa \= aaa \= aaa \= aaa \= aaa \= aaa \= aaa \= aaa \= aaa \= aaa \= aaa \= aaa \kill
\>  (PD-1)\> \> For  $i \in 1..m$, $j \in 1..m$: $M_{ij} \Rightarrow u_i \land v_j$ \\
\>  (PD-2)\> \> $\le_1(u_1,u_2,\ldots,u_m)$ \\
\>   (PD-3)\> \> $\le_1(v_1,v_2,\ldots,v_m)$ 
\end{tabbing}
The number of clauses generated by PD is characterized by $f(n) = 2n+ 2f(\sqrt n)$, and the number of new variables is characterized by $g(n) = 2\sqrt n + 2g(\sqrt n)$.

The \textit{\textbf{ sequential-counter}} (SC) encoding \cite{Sinz05} for $\le_k(x_1,x_2,\ldots,x_n)$ successively counts the number of $x_i$'s that are 1, in the following fashion: $c_1 = x_1$, $c_2 = c_1+x_2$, $\ldots$, $c_n = c_{n-1}+x_n$. Each count is a unary (base-1) number with $k$ bits: $c_i = $$<$$c_i^1c_i^2\ldots c_i^k$$>$, where $c_i^j = 0$ for $j \in i+1..k$, and $c_i^j = 1$ entails $c_i^{j-1} = 1$ for $j \in 2..k$. SC decomposes the constraint into the following:
  \begin{tabbing}
aaa \= aaa \= aaa \= aaa \= aaa \= aaa \= aaa \= aaa \= aaa \= aaa \= aaa \= aaa \= aaa \= aaa \kill
\> (SC-1)\> \>  For  $i \in 2..n-1$: $x_i \Rightarrow c_i^1$ \\
\> (SC-2)\> \>  For  $i \in 2..n-1$, $j \in 1..k$: $c_{i-1}^j \Rightarrow c_i^j$  \\
\> (SC-3)\> \>  For  $i \in 2..n-1$, $j \in 2..k$: $x_i \land c_{i-1}^{j-1} \Rightarrow c_i^j$ \\
\> (SC-4)\> \>  For  $i \in k+1..n$: $x_i \Rightarrow \neg c_{i-1}^k$ 
\end{tabbing}
  Constraint SC-1 ensures that if $x_i =1$ then the first bit of $c_i$ is 1. Constraint SC-2 ensures that monotonicity of addition: if the $j$th bit of $c_{i-1}$ is 1, then the $j$th bit of $c_i$ is also 1. Constraint SC-3 ensures that $c_i = c_{i-1}+1$ if $x_i = 1$. Constraint SC-4 ensures that no count exceeds $k$. The number of clauses and the number of new variables generated by SC are both $O(nk)$.

  The \textit{\textbf{ parallel-counter}} (PC) encoding \cite{Sinz05} decomposes $\le_k(x_1,x_2,\ldots,x_n)$  into the following:
  \begin{tabbing}
aaa \= aaa \= aaa \= aaa \= aaa \= aaa \= aaa \= aaa \= aaa \= aaa \= aaa \= aaa \= aaa \= aaa \kill
\> (PC-1)\> \>  $t\ =\ $\texttt{sum}$(x_1,x_2,\ldots,x_n)$ \\
\> (PC-2)\> \>  $t\ \le\ k$
  \end{tabbing}
  The function \texttt{sum}$(x_1,x_2,\ldots,x_n)$ returns a binary counter that represents the number of 1's in $x_i$'s with $m = \round{\log_2(k)+0.5}$ bits. Constraint PC-2 is enforced using a binary comparator, which compares the sum and the binary representation of $k$ from the highest bit to the lowest bit. The function \texttt{sum}$(x_1,x_2,\ldots,x_n)$ is defined as follows: if $n = 2$, then the two variables $x_1$ and $x_2$ are added using a half adder; if $n = 3$, then the three variables are added using a full adder; otherwise, the variables are split into two halves, each is summed recursively, and the results are added using a ripple-carry adder. Since the ultimate goal is to ensure that the final count never exceeds k, the binary counter only needs enough bits to count up to $k$, and incomplete adders that only propagate 1's are used. A half adder is implemented with 3 clauses, and a full adder is implemented with 7 clauses. The number of clauses and the number of variables generated by PC are both $O(mn)$.

  The \textit{\textbf{ binary-adder}} (BA) encoding implemented in Picat \cite{ZhouK17} treats $\le_k(x_1,x_2,\ldots,x_n)$ as a linear arithmetic constraint for $k > 1$. It repeatedly combines variables that have the smallest domains into a new variable until the constraint is reduced to the primitive form $t\ \le\ k$. It uses log encoding for all the newly introduced variables, uses complete binary adders for addition constraints, and enforces the constraint $t_i\ \le\ k$ on all new variables. Unlike incomplete adders used in PC that only propagate 1's, complete adders used in BA propagate both 1's and 0's. A complete half adder is implemented with 7 clauses, and a complete full adder is implemented with 10 clauses. 

  \section{EXPERIMENTAL RESULTS}
  Experiments were conducted to evaluate the encodings using the \textit{pigeonhole} problem as the benchmark. Given $P$ pigeons and $H$ holes, each of which can hold $K$ pigeons, the goal of the problem is to put the pigeons into the holes such that every pigeon is assigned a hole and no more than $K$ pigeons are put into any hole. The benchmark uses a $P\times H$ matrix of variables, $B$, and enforces the following constraints:
\begin{tabbing}
  aaa \=   aaa \=  aaa \= aaa \= aaa \= aaa \= aaa \= aaa \= aaa \= aaa \= aaa \kill
  \> (C-1)\> \> For $p \in 1..P$: $\sum_{h=1}^{H}B_{ph} = 1$  \\
  \\
  \> (C-2)\> \> For $h \in 1..H$: $\sum_{p=1}^{P}B_{ph} \le K$
\end{tabbing}
Constraint C-1 ensures that every pigeon is assigned a hole, and constraint C-2 ensures that every hole receives at most $K$ pigeons. Obviously, if $P > H\times K$, then the constraints are unsatisfiable.

The Maple SAT solver \footnote{\url{shorturl.at/joCIK}} was used in the experiments. All the CPU times reported below were measured on Linux Ubuntu with an Intel i7 3.30GHz CPU and 32G RAM.

  Table \ref{tab:amo} compares the encodings on CPU time (the sum of the compilation and solving times) for the at-most-one constraint. The first five instances are small and unsatisfiable, and the remaining five instances are large and satisfiable. While there are no significant differences in the speed for the small instances, PD is outstandingly fast for the large instances.
  
\begin{table}
  \begin{scriptsize}
    \begin{center}
      \caption{\label{tab:amo}A comparison of encodings for $\le_1$ (CPU time, seconds)}
      \begin{tabular}{|c|r|r|r|r|} \hline
        P-H-K  & BS  & PC  & PD  & SC \\ \hline
        12-11-1 & 4  & 9  & 9  & 5 \\
        13-12-1 & 13  & 15  & 8  & 10 \\
        14-13-1 & 75  & 28  & 46  & 16 \\
        15-14-1 & 123  & 47  & 119  & 42 \\
        16-15-1 & 511  & 229  & 549  & 216 \\
        100-100-1 & 901  & 1  & 1  & 76 \\
        200-200-1 & $>$1200  & 4  & 1  & $>$1200 \\
        300-300-1 & $>$1200  & 25  & 2  & $>$1200 \\
        400-400-1 & $>$1200  & $>$1200  & 6  & $>$1200 \\
        500-500-1 & $>$1200  & $>$1200  & 10  & $>$1200 \\ \hline
      \end{tabular}
    \end{center}
  \end{scriptsize}
\end{table}

Table \ref{tab:amk} compares the BA, PC, and SC encodings on CPU time for the at-most-k constraint (k $>$ 1). For constraint C-1, PD was used in all the runs, so this experiment evaluates the encodings for constraint C-2. BA is significantly faster than PC and SC, especially on the large instances. 

\begin{table}
  \begin{scriptsize}
    \begin{center}
      \caption{\label{tab:amk}A comparison of encodings for $\le_k$ (CPU time, seconds)}
      \begin{tabular}{|c|r|r|r|} \hline
        P-H-K  & BA  & PC  & SC \\ \hline
        19-9-2 & 47  & 50  & 53 \\
        21-5-4 & 4  & 9  & 4 \\
        22-7-3 & 49  & 177  & 51 \\
        25-6-4 & 50  & 459  & 325 \\
        26-5-5 & 9  & 41  & 501 \\
        100-20-5 & 7  & 49  & 45 \\
        200-40-5 & 19  & $>$1200  & $>$1200 \\
        300-60-5 & 41  & $>$1200  & $>$1200 \\
        400-80-5 & 72  & $>$1200  & $>$1200 \\
        500-100-5 & 105  & $>$1200  & $>$1200 \\ \hline
      \end{tabular}
    \end{center}
  \end{scriptsize}
\end{table}

\section{CONCLUSION}
This paper has presented an experiment that shows astounding performance of BA for the at-most-k (k $>$ 1) constraint in comparison with SC and PC. The major differences between BA and PC are that, BA uses complete adders and enforces cardinality on all new variables, while PC uses incomplete adders and only enforces cardinality on the final variable. The result entails that the clauses added by BA, while redundant, are helpful in enhancing the speed.

\end{document}